\documentclass{elsart}
\usepackage{epsfig}
\usepackage{lscape}
\usepackage{rotating}
\usepackage{array}

\usepackage{graphics}

\begin{document}

\begin{frontmatter}

\title{Test of a modified BCS theory performance in the Picket Fence Model}

\author[TUD,JINR]{V.Yu. Ponomarev}
\author[JINR]{and A.I. Vdovin}

\address[TUD]{Institut f\"ur Kernphysik,
Technische Universit\"at Darmstadt,
D--64289  Darmstadt, Germany}
\address[JINR]{Bogoliubov Laboratory of Theoretical Physics,
Joint Institute for Nuclear Research,
141980 Dubna, Russia}

\begin{abstract}

Analyses of a modified BCS (MBCS) theory performance at finite
temperatures in the Picket Fence Model (PFM) for light and heavy
systems is presented. Both symmetric, $\Omega=N$ ($N$ particles on
$\Omega$ twice-degenerate levels), and asymmetric, 
$\Omega\neq N$, versions of the PFM are considered. 
The quantities known exactly from particle-hole
symmetry of the $\Omega=N$ PFM are calculated.
Starting from very low temperatures, these quantities are found in 
dramatic deviation from the exact values in MBCS results.
Consequences of the MBCS prediction 
that heating generates a thermal constituent of the pairing gap,
are discussed.
Thermodynamical inconsistency of the MBCS is also addressed.

\end{abstract}

\end{frontmatter}

\section{Introduction\label{s1}}

A modified BCS (MBCS) theory for treating pairing correlations in
atomic nuclei at finite temperatures~\cite{DZ01,DA03a} has been
recently tested~\cite{D07} in the Picket Fence Model (PFM) in which 
$N$ particles are distributed over $\Omega$ twice-degenerate levels. 
The PFM with $N=\Omega$ is usually considered in the literature.

The MBCS predicts a smooth decreasing behavior for a pairing gap
as temperature $T$ increases up to some $T_M$ when the theory
suddenly breaks down. It was reported~\cite{D07,DA06} that adding
one extra level ($\Omega = N+1$) extends the MBCS applicability to
much higher temperatures. As is already established~\cite{PV06},
the $\Omega = N+1$ case of the PFM is the only example of an
exceptional MBCS pairing gap behavior and thus, it could be hardly
considered as a typical one. Accordingly, we find it necessary to
provide the reader with other $\Omega\neq N$ PFM examples not
available in \cite{D07}. This is done in Sec.~\ref{s2}.

In Sec.~\ref{s3}, the MBCS prediction that not only the pairing
force but also the heating itself generates the pairing gap, is
discussed.

In Sec.~\ref{s4}, we address the key question of the article: whether
the MBCS is a reliable theory in the temperature domain where its
pairing gap looks reasonable at first glance. For that, we examine
some quantities which are known exactly in the $N=\Omega$ PFM
because of symmetry. We show that in the MBCS predictions,
these quantities dramatically deviate from their exact values
starting from very low temperatures.

Thermodynamical inconsistency of the MBCS is discussed in Sec.~\ref{s5}.

\section{MBCS pairing gap in PFM systems with $\Omega=N$ and
$\Omega\neq N$ \label{s2}}

The PFM or Richardson model is widely used as a test model for the
pairing problem. It is the pairing Hamiltonian applied to a system
of $N$ fermions distributed over $\Omega$ equidistant levels. All
levels are twice degenerate for the spin up and down. The levels
below (above) the Fermi surface will be referred to as holes
(particles) and labeled by ``$-i$'' (``$i$''). Their single
particle energies are $\varepsilon_{-i} = (0.5 - i)$~MeV for holes
and $\varepsilon_{i} = (-0.5 + i)$~MeV for particles, where $i =
1,~2,~,\ldots$ (i.e. $\varepsilon_{i} = - \varepsilon_{-i}$). In
all calculations presented below (except for the ones in Fig.~\ref{fig6}a),
the pairing strength parameter $G$ is adjusted so that the pairing
gap $\Delta$ equals 1~MeV at zero temperature.

In addition to the fact that exact solution of the PFM is possible
if $N$ and $\Omega$ are not big, the PFM with $N=\Omega$ (to be
referred to as the conventional PFM) possesses internal
particle-hole symmetry. This means that at any temperature:

a) the energy of the Fermi surface $E_F$ equals exactly 0~MeV:
\begin{equation}
E_F \equiv 0~;
\label{eq1}
\end{equation}

b) the quasiparticle energies $E_i$ for particles and holes should be
degenerate:
\begin{equation}
E_i = \sqrt{(\varepsilon_i - E_F)^2 + \Delta^2} \equiv
 \sqrt{(\varepsilon_{-i} - E_F)^2 + \Delta^2} = E_{-i} 
\label{eq2}
\end{equation}
because of (\ref{eq1});

c) the particle occupation probabilities (Bogoliubov $u_i$ and $v_i$
coefficients) are related as:
\begin{equation}
u_i = \sqrt{\frac{1}{2}
\left( 1 + \frac{\varepsilon_i - E_F}{E_i}\right)}
\equiv
\sqrt{\frac{1}{2}
\left( 1 - \frac{\varepsilon_{-i} - E_F}{E_{-i}}\right)}
=
v_{-i}
\label{eq3}
\end{equation}
because of (\ref{eq1},\ref{eq2});

d) the thermal quasiparticle occupation numbers
$n_i = 1 / (1+ \exp{(E_i/T)})$
and quasiparticle-number fluctuations 
$\delta N_i = \sqrt{n_i (1-n_i)}$
should be equal for particles and holes with the same $i$:
\begin{equation}
n_i \equiv n_{-i}~~~~~{\rm and}~~~~~\delta N_i \equiv \delta N_{-i}~
\label{eq4}
\end{equation}
because of (\ref{eq2}).

Of course, an asymmetric version of the PFM with $\Omega = N + k$
(where $k = -N/2 + 1,\ldots, -1,$ $1,\ldots,$ $\infty$)
may be considered as well but Eqs.~(\ref{eq1}-\ref{eq4}) are not
valid for it.

The first test of the MBCS performance in the conventional PFM
with $N=\Omega=10$ revealed that at $T \approx 1.75$~MeV the system 
undertakes a phase transition which manifests itself as a sharp 
simultaneous increase in the pairing gap, a sharp decrease of the system 
energy and a discontinuity in the specific heat $C_V$ (this phase transition 
was defined as a superfluid -- super-superfluid phase 
transition)~\cite{PV05}. 
This critical temperature was denoted as $T_M$~\cite{D07,DA06}.
It was found that $T_M$ linearly increases with the number of particles
$N$ in the conventional PFM.

\begin{figure}
\begin{center}
\epsfig{file=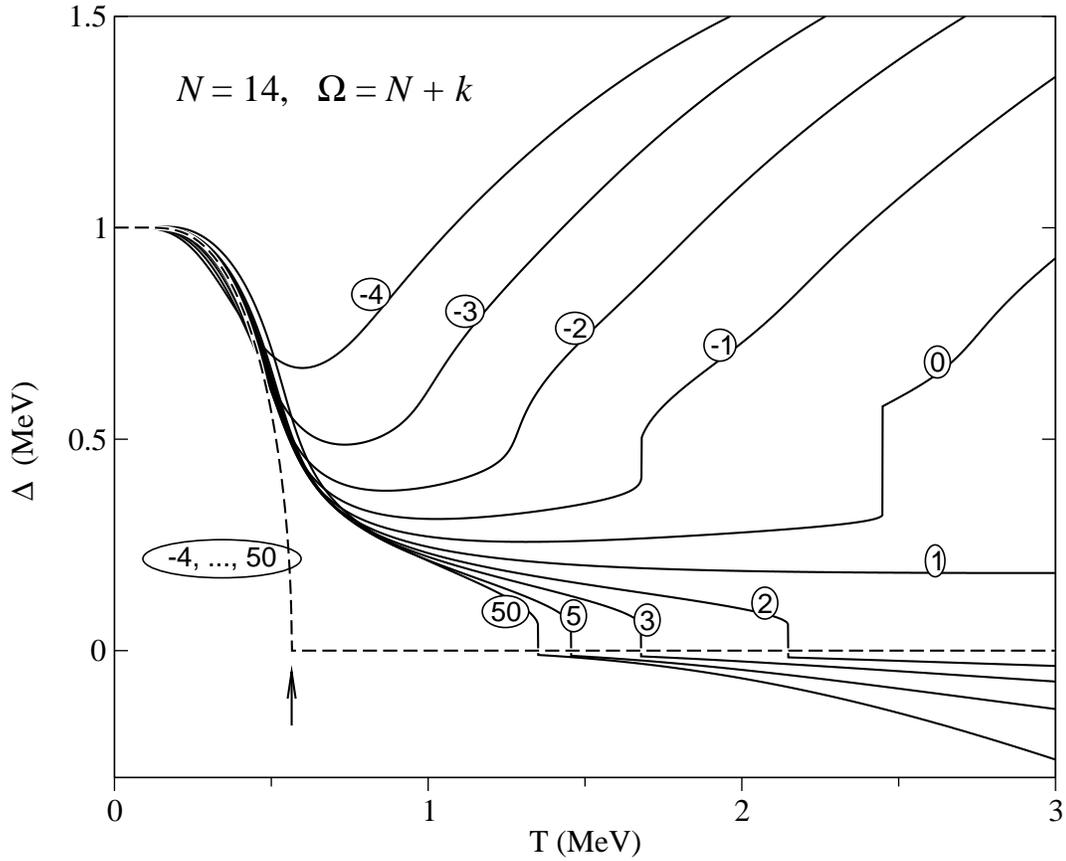,width=14.cm,angle=0}
\end{center}
\caption{MBCS pairing gap $\Delta$ for the system of $N = 14$
particles on $\Omega = N + k$ levels (solid curves). The BCS
pairing gap is plotted by the dashed curve. The $k$ values are
indicated in an oval at each curve. The conventional critical
temperature $T_c$ is shown in this and all other figures by the
vertical arrow.} \label{fig1}
\end{figure}

It was also reported that enlarging the space by one more level, 
$\Omega = N + 1$, restores the MBCS applicability to much higher 
temperatures even for $N \leq 14$ systems~\cite{D07,DA06}. 
Fig.~\ref{fig1} plots the MBCS pairing gap for $N = 14$ particles 
and $\Omega = N + k$ levels with $k$ changing from $-4$ to 50 (solid 
curves).
Indeed, the MBCS gap above $T_c$ is small and almost constant up to 
$T_M = 7.2$~MeV when suggested example of extended configuration spaces,
$k=1$, is considered.
However, the MBCS theory breaks down again at rather modest temperature 
$T_M=2.15$~MeV with adding one more extra level, $k=2$, etc.
Thus, the extension of the configuration space beyond $k=1$ makes the MBCS
results inappropriate again.

One immediately notices from Fig.~\ref{fig1} that the MBCS
theory predicts two typical scenarios for the system evolution with
heating.
As temperature increases, the system undertakes either

a) a superfluid -- super-superfluid phase transition
(examples with $k \leq 0$ in Fig.~\ref{fig1})

or

b) a phase transition from a superfluid phase with a
positive gap to another superfluid phase but with a negative gap
(examples with $k > 1$ in Fig.~\ref{fig1}),

instead of a superfluid -- normal phase transition of the
conventional BCS.
Corresponding BCS pairing gap behavior in the same systems is shown in
Fig.~\ref{fig1} by the dashed curve. It is impossible to visually
distinguish the results when $k$ value changes from $-4$ to 50.

We have performed additional calculations for the asymmetric PFM
with $N$ changing from 6 to 100. The results look very similar to
the ones in Fig.~\ref{fig1}. In all these examples, there exists
only a single case $k = 1$ with abnormally large $T_M$
which grows almost linearly with $N$.
For all other $k$ values, the phase transitions of unknown types take
place at much lower $T_M$ which becomes smaller and smaller as $k$
differs from $1$ larger and larger.

However, only the case $\Omega = N + 1$ has been selected for
presentation in \cite{D07,DA06} as an example of an extended
configuration space ($\Omega > N $).

\section{MBCS pairing gap: quantal and thermal constituents
\label{s3}}

Although the MBCS performance has been already discussed in
several articles, a clear answer to the question what is
responsible for phase transitions of unknown types predicted by
this theory is not yet established. 
According to the MBCS founders, there exists a criterion of the MBCS
applicability, according to which the line shape of the
quasiparticle-number fluctuations $\delta N_i$ should be
symmetric~\cite{D07,DA06}.
And the theory reaches the limit of its applicability at $T_M$.
The physical origin of this criterion is unclear. 
The physical spectra are never symmetric, at least in nuclear physics.
On the other hand, the single particle spectrum of the
conventional PFM is ideally symmetric with respect to a chemical
potential, see Eqs.~(\ref{eq2},\ref{eq4}), but the MBCS itself
breaks down this symmetry (see below).

\begin{figure}
\begin{center}
\epsfig{file=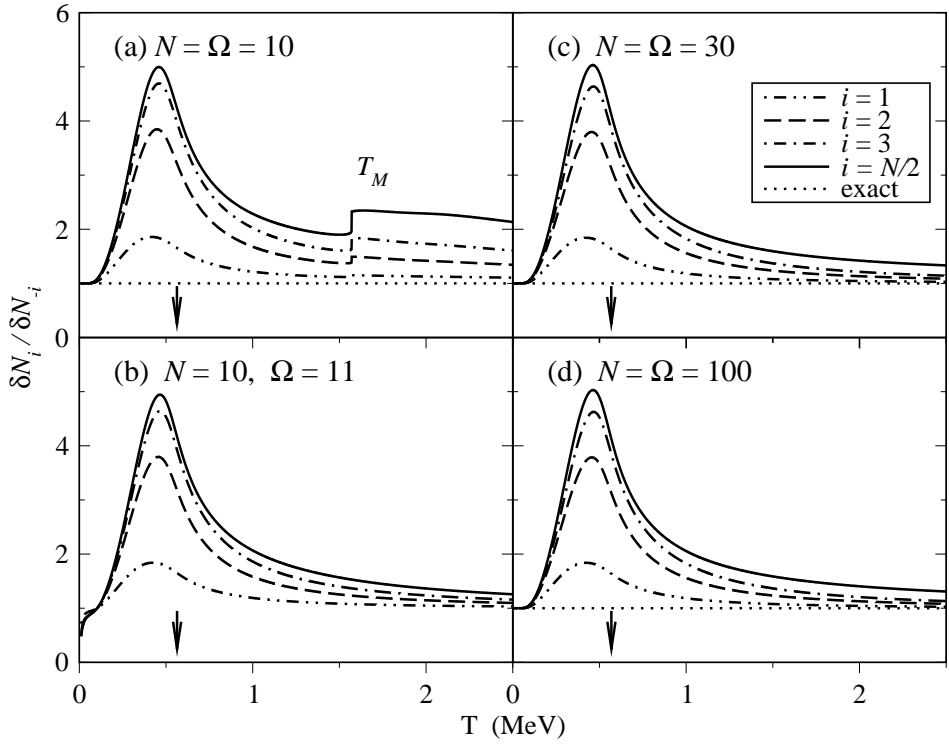,width=12.5cm,angle=0}
\end{center}
\caption{MBCS predictions for the ratios of quasiparticle-number
fluctuations $\delta N_i/\delta N_{-i}$ in different $(N, \Omega)$
systems. Exact result $\delta N_i/\delta N_{-i} \equiv 1$ in Figs.
\ref{fig2}(a,c,d) is shown by the dotted line.} \label{fig2}
\end{figure}

The reader of \cite{D07} is suggested to notice from Fig.~2 that the
line shape becomes rather asymmetric approaching $T_M$, thus
violating the above-mentioned criterion of the MBCS applicability.
We find it difficult to judge of the symmetry of lines in
Figs.~2(a-e). On the other hand, it is very easy to quantify the
results in this figure by plotting the $\delta N_i/\delta N_{-i}$
ratios as a function of temperature for different $i$. This is
done in our Fig.~\ref{fig2} for different systems.

Indeed, the asymmetry increases above $T_M$ in the $N=\Omega=10$
system, Fig.~\ref{fig2}(a), while this does not in more heavy symmetric 
$N=\Omega$, Figs.~\ref{fig2}(c,d), or 
selected asymmetric $\Omega=N+1$, Figs.~\ref{fig2}(b), systems. 
However, it is difficult to miss that this effect is marginal
compared to asymmetry in all systems (symmetric, asymmetric, light, 
and heavy) at much lower $T$ with the maximum around $T_c$. 
In other words, the most severe violation
of the suggested criterion of the MBCS applicability takes place
around the critical temperature of the conventional BCS and
not at $T_M$ where the MBCS breaks down, as stated in~\cite{D07,DA06}. 
This conclusion does not depend on details of the PFM system being
considered.

Since the suggested criterion does not help to understand what causes
the theory breaking down, we continue our analysis.
Let us read once again the paragraph containing Eq.~(26) in Ref.~\cite{D07}
that an important feature of the MBCS theory is that the MBCS gap is the
sum of a quantal part (which looks the same as in the conventional
BCS) and a thermal part $\delta \Delta$:
\begin{eqnarray}
\delta \Delta_h &= G\sum_{i}^{\rm holes}
(v_i^2-u_i^2)\delta N_i
&~~~{\rm for~holes}
\label{eq5}
 \\
\delta \Delta_p &= G\sum_{i}^{\rm particles}
(v_i^2-u_i^2)\delta N_i
&~~~{\rm for~particles}~.
\label{eq6}
\end{eqnarray}

\begin{figure}
\begin{center}
\epsfig{file=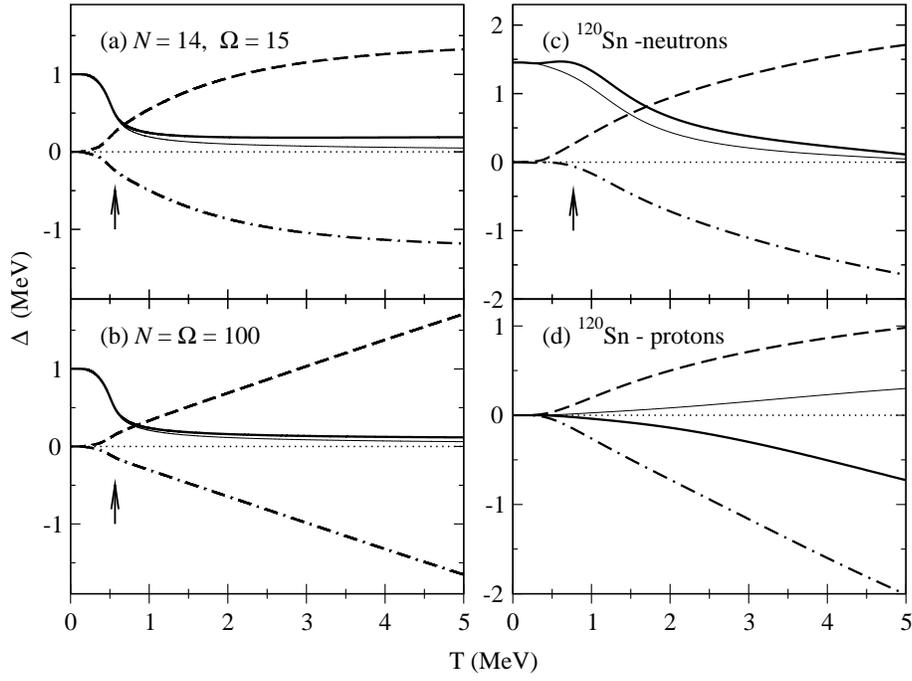,width=12.cm,angle=0}
\end{center}
\caption{MBCS pairing gap (thick solid curve) for the PFM systems with
(a) $N=14$, $\Omega =15$ and (b) $N=\Omega =100$ and in calculation
with a realistic single particle spectrum for $^{120}$Sn: (c) neutron
and (d) proton systems.
This gap is the sum of a quantal (particles plus holes -- thin
solid curve) and thermal (holes -- dashed curve, particles --
dot-dashed curve) parts.}
\label{fig3}
\end{figure}

Fig.~\ref{fig3} presents the MBCS gap for (a) light and (b)
heavy PFM systems by the thick solid curve. The quantal part of it
is shown by the thin solid curve, it becomes very small soon above
$T_c$. The thermal part for holes (particles) is plotted by the
dashed (dot-dashed) curve. 
The quantity $\delta \Delta_h$ is always positive ($v_i > u_i$) and 
the quantity $\delta \Delta_p$ is always negative ($u_i > v_i$); their
absolute values increase with temperature.

The physics of pairing suggested by the conventional BCS is very simple
and transparent.
Pairing is generated by the pairing force.
As temperature increases the thermal scattering of
nucleons becomes stronger and stronger and finally destroys the pairing
at $T_c$.

The MBCS suggests another physics, according to which the heating
itself generates a thermal constituent of the pairing gap: positive for
holes and negative for particles.
The heavier is the system, the stronger thermal
pairing gap may be generated. 
A similar phenomenon takes place in calculation with realistic
single particle spectra (see Figs.~\ref{fig3}(c) and (d) where the
pairing gap behavior in  $^{120}$Sn is presented for neutrons and
protons, respectively).

Notice, when the pairing strength is weak to generate pairing at zero 
temperature, the MBCS predicts that the heating develops the pairing 
gap at finite $T$, as is obvious from Eqs.~(\ref{eq5}-\ref{eq6}). 
An example of the pairing induced by heating alone in a magic nuclear 
system is shown in Fig.~\ref{fig3}(d).
One should not be surprised that the resulting pairing gap is negative
in this example.
Fig.~\ref{fig3} clearly shows that as temperature increases, the MBCS gap 
receives main contribution from $\delta \Delta_h$ and $\delta \Delta_p$ 
terms and their sum can be equally positive (less particle levels,
$k<0$ in Fig.~\ref{fig1}) or negative (more particle levels, $k>1$ in 
Fig.~\ref{fig1}).

The main goal of the MBCS theory is to mimic the thermal behavior
of the pairing gap of a macroscopic treatment \cite{M72} that the
normal -- superfluid phase transition is washed out and soon above
$T_c$ the gap remains rather small but positively finite. However,
Fig.~\ref{fig3} shows that even technically, it is very difficult
to achieve the desirable goal with the MBCS equations: it is
necessary that two almost linearly growing functions $\delta \Delta_h$ and
$-\delta \Delta_p$ almost cancel each other with a high
accuracy in a large temperature interval\footnote{Independently
from the physical content of the thermal gap in the MBCS, one finds 
from Eqs.~(\ref{eq3}-\ref{eq6}) that
$\delta \Delta_h \equiv -\delta \Delta_p$ for the conventional PFM
and the cancellation
between these two terms should be exact at any $T$. This does not
happen in MBCS calculations because the MBCS theory violates
Eqs.~(\ref{eq3},\ref{eq4}), see below.}. Of course, the final
result is very sensitive to tiny details of a single particle
spectrum employed. If a spectrum with desirable properties is
occasionally found, the theory breaks down anyway with a step
aside from it. Fig.~\ref{fig1} clearly demonstrates it.

\section{On range of the MBCS validity\label{s4}}

As is established, the MBCS theory is capable of generating the
pairing gap behavior which looks reasonable at first glance 
below $T_M$. It is also established that $T_M$ grows
linearly with the mass of the system and it is possible
to find a single exceptional case of abnormally large $T_M$. Does
this mean that the MBCS theory should be considered as reliable
for $T<T_M$ and recommended for applications?

To answer this question we return to the conventional PFM. Due to
its internal particle - hole symmetry, the accuracy of the MBCS
predictions for the theory variables can be easily examined even
without having exact solutions. 

Verification of Eq.~(\ref{eq4}) for the $\delta N_i$ quantities in
Figs.~\ref{fig2}(a,c,d) reveals that this property of the system
is enormously violated in the MBCS predictions.
The behavior of the ratios $n_i/n_{-i}$ is very similar to the one of
$\delta N_i / \delta N_{-i}$ in Fig.~\ref{fig2}. Deviation from
the exact result $n_i/n_{-i} \equiv 1$ reaches a few thousand per
cent near $T_c$.

\begin{figure}
\begin{center}
\epsfig{file=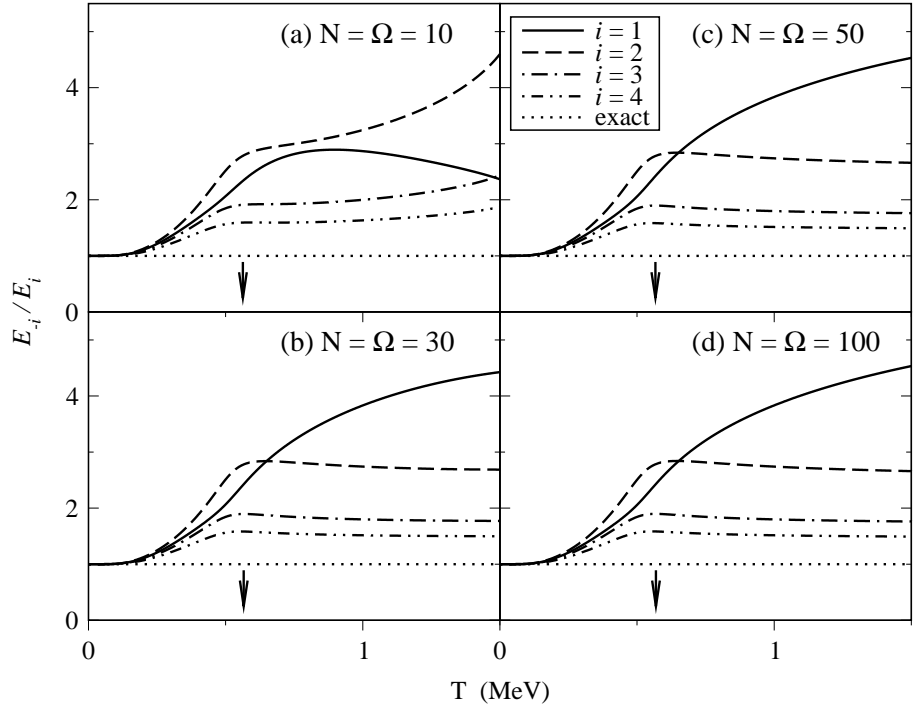,width=12.cm,angle=0}
\end{center}
\caption{MBCS predictions for the ratios of quasiparticle energies
$(E_{-i}/E_{i})$ in different $N=\Omega$ systems. Exact results
$(E_{-i}/E_{i}) \equiv 1$ are shown by the dotted line.}
\label{fig4}
\end{figure}

\begin{figure}
\begin{center}
\epsfig{file=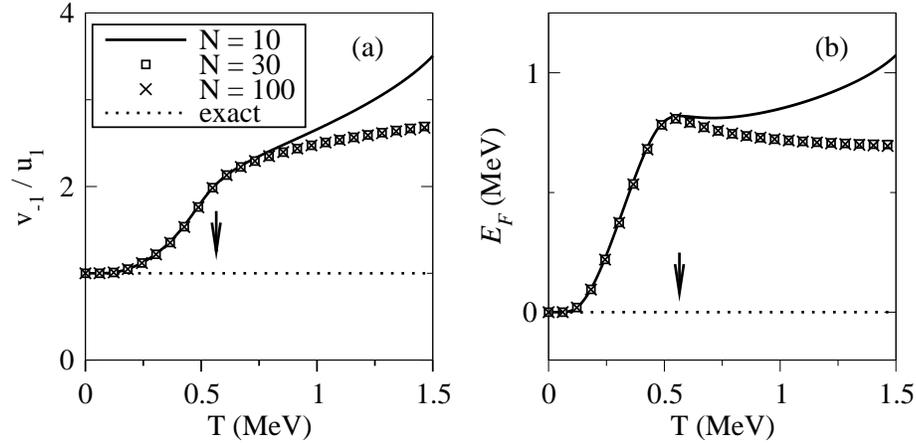,width=12.cm,angle=0}
\end{center}
\caption{MBCS predictions for (a) ratios $(v_{-1}/u_1)$ and
(b) the energy of the Fermi level $E_F$
 in $N=\Omega =$ 10, 30, and 100 systems.
Exact results (a) $(v_{-1}/u_1) \equiv 1$ and (b) $E_F \equiv 0$
for these systems are shown by the dotted line.} \label{fig5}
\end{figure}

The accuracy of the MBCS predictions for the quasiparticle energies, 
Eq.~(\ref{eq2}), and for the particle occupation probability, 
Eq.~(\ref{eq3}), with $i = 1$ is examined in
Fig.~\ref{fig4} and Fig.~\ref{fig5}(a), respectively.
Fig.~\ref{fig5}(b) shows that the MBCS theory does
not keep the energy of the Fermi surface at zero energy, as it
should be, Eq.~(\ref{eq1}).

The results in Figs.~\ref{fig2}(a,c,d), \ref{fig4}, and \ref{fig5}
obviously demonstrate that as soon as heating starts to play a
role, the MBCS predictions are irrelevant to a system
which the theory aims to describe, because genuine
properties of the system are severely violated in these predictions.
It does not matter whether a light or heavy systems is considered,
the results are almost independent on $N$.

All-in-all, and this is the main conclusion of the article:
although the MBCS theory generates a pairing gap which
looks reasonable in some temperature interval $[0,T_M]$, it is
clear from the analysis of other quantities that the theory
predictions have nothing to do with the properties of a system under
discussion even in this temperature interval.

\section{Thermodynamic inconsistency of the MBCS theory\label{s5}}

We remind the reader that expressions for the MBCS theory were copied from
zero-temperature BCS expressions by straightforward replacing the BCS
$\{u_j,v_j\}$ coefficients with the MBCS $\{\bar{u}_j,\bar{v}_j\}$ 
coefficients.
It has been already pointed out~\cite{PV05} that a method of
mechanical copying of equations from one theory into another one has
absolutely no grounds because the properties of $\{u_j,v_j\}$ and
$\{\bar{u}_j,\bar{v}_j\}$ coefficients are simply different.
E.g., the thermal part of the pairing gap $\delta\Delta$, discussed
in Sec.~\ref{s3}, has appeared in the MBCS theory as a result of such
an uncontrolled copying.

\begin{figure}
\begin{center}
\epsfig{file=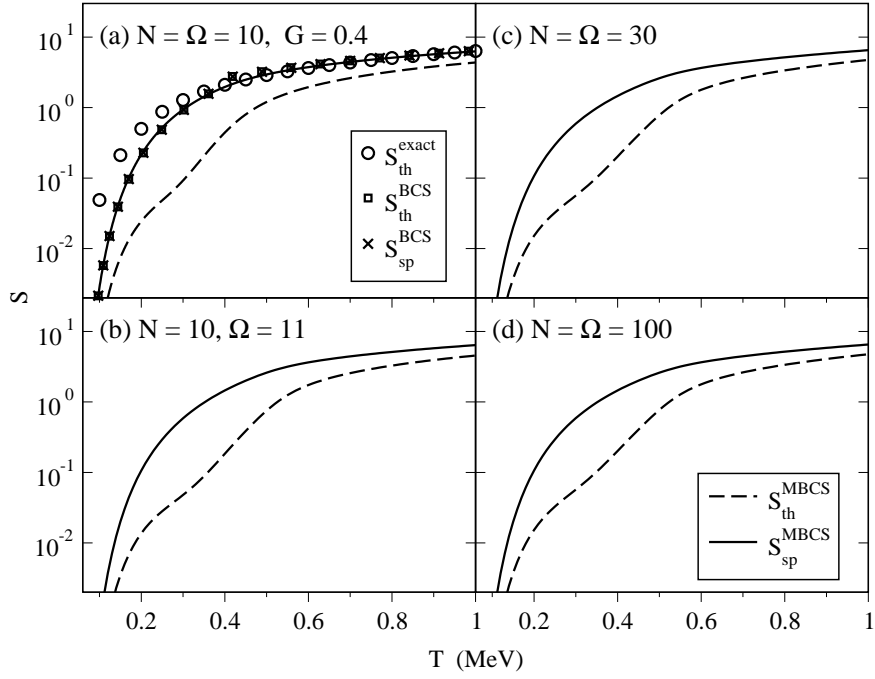,width=11.5cm,angle=0}
\end{center}
\caption{MBCS predictions for the system entropy calculated as
$S_{\rm th}$, Eq.~(\ref{Sth}), and $S_{\rm sp}$, Eq.~(\ref{Ssp}) in different
light and heavy, symmetric and asymmetric PFM systems. Notice the
logarithmic $y$ scale.}
\label{fig6}
\end{figure}

It is very natural to verify whether expressions for thermodynamical
observables obtained by such a method are thermodynamically consistent
in the MBCS theory.
Fig.~\ref{fig6} compares the system entropy which
is calculated according its thermodynamical definition (dashed lines):
\begin{equation}
S_{\rm th} = \int_0^T \, \frac{1}{\tau} \,\cdot \,
\frac{\partial  E_{tot}}{\partial \tau} \,d\tau
\label{Sth}
\end{equation}
and within a statistical approach (solid lines):
\begin{equation}
S_{\rm sp} = - \sum_j \, (2j+1)\, \left [ \, n_j \, {\rm ln} \,
n_j + (1-n_j) \,
{\rm ln} \, (1-n_j) \right ]~~.
\label{Ssp}
\end{equation}
The latter one is also called the single particle entropy in
Refs.~\cite{Zel,Kot} or the quasiparticle entropy in Ref.~\cite{D07a}.

Fig.~\ref{fig6} shows that $S_{\rm th}$ and $S_{\rm sp}$ being considered 
separately are amazingly close in different PFM systems for $T\leq1$~MeV
but they are dramatically different one from another in the MBCS predictions.
The latter fact has been already addressed~\cite{PV06} when 
$S_{\rm th}^{\rm MBCS}$ and $S_{\rm sp}^{\rm MBCS}$ for the neutron system 
of $^{120}$Sn have been found different by two orders of magnitude.

Different entropy-like quantities in nuclear physics have been considered
in \cite{Zel,Kot}.
It has been found that ``the thermodynamic entropy $\ldots$, the information
entropy $\ldots$ and the single-particle entropy $\ldots$, all
coincide for strong enough interaction but only in the presence
of a mean field'' \cite{Kot}.
It is not possible to accept interpretation of these results in 
Ref.~\cite{D07a} as that $S_{\rm th}$ and $S_{\rm sp}$ ``are nearly the 
same only for noninteracting particles''.
Notice, correspondence between $S_{\rm th}$ and $S_{\rm sp}$ in a mean field
plus residual forces (Figs.~56(IIa, IIc) in \cite{Zel}) is  
definitely not worse than in an almost pure mean field approach 
(Figs.~56(Ia, Ic) in \cite{Zel}) .

Fig.~\ref{fig6}(a) also presents the quantities $S_{\rm th}^{\rm BCS}$ 
and $S_{\rm sp}^{\rm BCS}$ calculated within the conventional BCS theory 
and $S_{\rm th}^{\rm exact}$  obtained from the exact solution of the PFM 
for the $N=\Omega=10$ system.
One notices that for $T>100$~keV:
\begin{equation}
S_{\rm th}^{\rm exact} \approx S_{\rm th}^{\rm BCS} \approx 
S_{\rm sp}^{\rm BCS} \approx S_{\rm sp}^{\rm MBCS} >>
S_{\rm th}^{\rm MBCS}~~.
\label{comp}
\end{equation}

An attempt to compare $S_{\rm th}^{\rm MBCS}$ and $S_{\rm sp}^{\rm
MBCS}$ with the entropy calculated from the exact solution of the
PFM has been made in Fig.~6 of Ref.~\cite{D07a} for the
$N=10$, $\Omega =11$ system. 
First of all, the calculation of $S_{\rm sp}^{\rm exact}$ 
(thick solid line in Fig.~6(a) of Ref.~\cite{D07a}) is definitely
not correct.
The author of \cite{D07a} claims that $S_{\rm sp}^{\rm exact}$
does not vanish at $T=0$  because occupation probabilities $f_h < 1$ 
and $f_p >0$.
The problem with the third law of thermodynamics\footnote{``at absolute 
zero, any part of the body must be in a definite quantum state - namely 
the ground state $\ldots$ the entropy of the body - the logarithm of its 
statistical weight - is equal to zero'' (see $\S$23, page 66 in \cite{LL}).}
in this calculation is caused by confusion of interacting particles with  
noninteracting ``quasiparticles'':
particle levels are not eigen states of the pairing Hamiltonian, 
their occupation numbers do not obey Fermi-Dirac distribution
and because of that, they cannot be used in Eq.~(\ref{Ssp}) which
represents the free Fermi-gas combinatorics.
Incorrectly calculated $S_{\rm sp}^{\rm exact}$ is also published in 
Fig.~8 of \cite{D07}.

One notices that Fig.~6(b) of Ref.~\cite{D07a} lacks comparison between
$S_{\rm th}^{\rm MBCS}$ and $S_{\rm th}^{\rm exact}$ quantities%
\footnote{What is plotted by thick solid line in this figure and makes
impression of agreement with $S_{\rm th}^{\rm MBCS}$ is not specified.}. 
The latter quantity can be easily calculated 
from $E_{tot}^{\rm exact}$ (thick solid line in Fig.~4(a) of 
Ref.~\cite{D07a}).
A level of expected disagreement between $S_{\rm th}^{\rm MBCS}$ and 
$S_{\rm th}^{\rm exact}$ quantities can be seen from our 
Fig.~\ref{fig6}(a) (compare dashed curve with circles).

Analysis of the MBCS predictions for the system entropy clearly indicate
a problem with the expression for the system energy which enters in
Eq.~(\ref{Sth}). How this expression was obtained, is mentioned in the
beginning of the section.
In Fig.~9 of Ref.~\cite{PV05} the system energy calculated from this
expression has been straightforwardly compared to the system energy
calculated as ${\rm <H>=Tr(HD)}$ where $D$ is a density operator and
$<\ldots>$ means averaging over the grand canonical ensemble.
Dramatic disagreement in two quantities representing the same physical
observable has been obtained indicating the same problem.

\section{Conclusions\label{s6}}

In this article we continue discussion on the validity of the MBCS
theory. 
The theory performance is examined within the PFM.
Our present goal is to allow the reader, who is not familiar 
with the previous discussion on the subject in 
Refs.~\cite{DA06,PV06,PV05,DA03b}, to better
understand the results and conclusions in Refs.~\cite{D07,D07a}.

We confirm that there exists a single example of the PFM (with the
number of levels $\Omega$ equal to the number of particles $N$
plus one) in which the MBCS produces the thermal behavior of the
pairing gap similar to the one of a macroscopic theory up to a
rather high temperature. In our opinion, this fact alone is not
sufficient to conclude on the theory performance without noticing
that in all other examples of the PFM with $\Omega \ne N$ the
theory predicts phase transitions of unknown types at a much lower
temperature. These other examples are neither shown nor discussed
in Ref.~\cite{D07}.

Although the MBCS formally yields a visually
acceptable pairing gap behavior in some temperature domain, the
physical content of its pairing gap is very dubious. The MBCS
predicts that heating generates a thermal constituent of the
pairing gap which becomes stronger and stronger with temperature
in contradiction with generally accepted understanding of the
pairing phenomenon in nuclei as a result of a specific
particle-particle interaction. We point out that this unphysical
constituent of the MBCS pairing gap is responsible for strong
sensitivity of the theory predictions to tiny details of single
particle spectra.

The conventional PFM with $\Omega = N$ possesses internal
particle-hole symmetry. Due to it, some quantities in the model
are known exactly without any calculations. 
We have demonstrated that the MBCS predictions deviate from exact results
by a few hundred per cent starting from very low temperatures.

We also point out that the MBCS is a thermodynamically inconsistent
theory.

Performing a systematic analysis of the MBCS performance in the
PFM we have failed to find a model system and/or temperature range
where the MBCS predictions are not questionable.

\section*{Acknowledgments}
The work was partially supported by the Deutsche
Forschungsgemeinschaft (SFB 634).

\end{document}